\documentclass[a4paper,showpacs,apl,twocolumn]{revtex4}%
\usepackage{amsfonts}
\usepackage{amsmath}
\usepackage{amssymb}
\usepackage{graphicx}%
\begin{document}
\title{Surface state photonic bandgap cavities}
\author{A. I. Rahachou and I. V. Zozoulenko}
\affiliation{Department of Science and Technology, Link\"{o}ping
University, 601 74, Norrk\"{o}ping, Sweden}
\date{\today}

\begin{abstract}
We propose and analyze a new type of a resonant high-$Q$ cavity
for lasing, sensing or filtering applications, which is based on a
surface states of a finite photonic crystal. We demonstrate that
such the cavity can have a $Q$ factor comparable with that one of
conventional photonic band-gap defect mode cavities. At the same
time, the distinguished feature of the surface mode cavity is that
it is situated directly at the surface of the photonic crystal.
This might open up new possibilities for design of novel photonic
devices and integration of photonic circuits.
\end{abstract}


\maketitle

Photonic crystals have attracted ever increasing attention in the
last decade due to their unique properties and possible
applications in future generation of optical and photonic devices
like light emitting diodes, delay lines, waveguides and lasers
\cite{joannopoulos,Sakoda}. The lasing effect has been
demonstrated for variety of photonic crystal structures including
bandgap defect mode lasers \cite{defect-mode}, distributed
feedback lasers \cite{feedback}, and bandedge lasers
\cite{bandedge}. High-$Q$ photonic bandgap cavities find their
potential application in all-optical networks and photonic chips
\cite{jon,Noda}. In this Letter we present a new type of a
photonic bandgap cavity utilizing a surface state residing on the
surface of a photonic crystal, which can be used for lasing,
sensing or filtering applications. We demonstrate that a weak
coupling of the resonant modes of the cavity with the outgoing
radiation results in a strongly enhanced intensity of an
electromagnetic field on the surface of the photonic crystal and
leads to the ultra-high cavity $Q$ factor.

Surface states reside at the interface between a photonic crystal
and open space, decaying both into the crystal and air
\cite{joannopoulos}. An unmodified surface of a semi-infinite
photonic crystal does not normally support the surface state. The
surface state appears in the band-gap of the photonic crystal when
the boundary of the photonic crystal is modified by, e.g.,
truncating the surface rods, shrinking or increasing their size,
or creating more complex surface geometry
\cite{joannopoulos,Mendieta,Zhang,Elson,we-PRB-2005}. It is
important to emphasize that the surface mode residing on the
infinitely long boundary of a semi-infinite crystal represents a
truly bound state with the infinite lifetime and $Q$ factor, and
consequently does not couple with the incoming/outgoing radiation.

In our recent work \cite{we-PRB-2005} we have demonstrated that
when the translational symmetry along the boundary of the
semi-infinite crystal is broken, the surface mode turns into a
resonant state with a finite lifetime, which can  be utilized for
lasing and sensing applications. The main purpose of the present
Letter is to propose and analyze a structure that can be used for
experimental realization of a high-$Q$ surface state cavity in the
bandgap photonic crystal. To identify the surface state resonant
modes and compute their quality factors, we apply a novel
computational approach based on the two-dimensional (2D) recursive
Green's function technique \cite{we-PRB-2005}. Using this
technique, the intensity distribution of the electromagnetic
radiation is computed in the frequency domain on the basis of the
the Green's function of the photonic structure that is calculated
recursively by adding slice by slice on the basis of Dyson's
equation. In order to account for the infinite extension of the
structure both into air and into the space occupied by the
photonic crystal we make use of the so-called "surface Green's
functions" that propagate the electromagnetic fields into (and
from) infinity.

\begin{figure}[htb!]
\begin{center}
\includegraphics[keepaspectratio,width=\columnwidth]{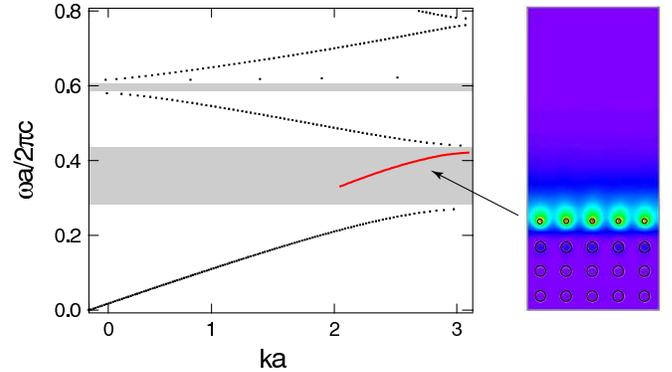}
\end{center}
\caption{(color online)  Dispersion relation  for the TM modes of
an infinite square-lattice photonic crystal in the $\Gamma X$
direction. The rod diameter $D=0.4a$, where $a$ is the lattice
constant; $\varepsilon = 8.9$. Shaded regions indicate the
bandgaps. The red line shows the dispersion relation for the
surface mode in a semi-infinite photonic crystal where the
diameter of the outermost surface rods $d=0.5D$. The inset depicts
the intensity of the $E_z$-component of the electromagnetic field
of the surface state. } \label{fig1-disp-rel}
\end{figure}
We study a semi-infinite square-lattice photonic crystal composed
of dielectric rods ($\varepsilon = 8.9$, diameter $D=0.4a$, $a$ is
a lattice constant) in the air background. This structure has a
fundamental bandgap for the TM-polarization in the interval $0.33
\leq \omega a / 2\pi c \leq 0.44$ \cite{joannopoulos}, see Fig
\ref{fig1-disp-rel}. To modify the surface we place rods of a
reduced diameter $d=0.5D$ on the infinitely long surface of a
semi-infinite photonic crystal. This configuration supports one
surface mode whose dispersion relation along the surface is
illustrated in Fig. \ref{fig1-disp-rel}. The inset to Fig.
\ref{fig1-disp-rel} shows the intensity of the $E_z$-component of
the electromagnetic field in the surface state. The field is
strongly localized at the surface rods and quickly decays both
into the bulk of the photonic crystal and into air.

\begin{figure}[htb!]
\begin{center}
\includegraphics[keepaspectratio,width=\columnwidth]{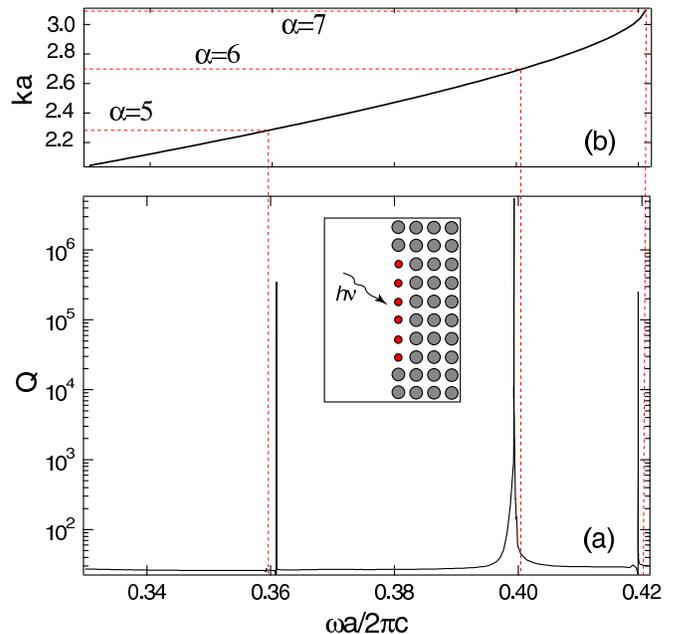}
\end{center}
\caption{(color online) (a) Frequency dependence of the $Q$ factor
of a surface state photonic bandgap cavity. Inset illustrates the
resonant cavity defined by the $N=6$ surface rods of the smaller
diameter $d=0.5D$ placed on the photonic crystal surface. (b) The
dispersion relation for the surface state for the semi-infinite
photonic crystal. The dashed lines indicate the expected resonant
wave vectors for the modes $\alpha=5,6,7$ given by Eq.
(\ref{resonant_frequency}) and corresponding expected resonant
frequencies. Parameters of the photonic crystal are the same as
for the lattice of Fig. \protect\ref{fig1-disp-rel}. The structure
is illuminated at the incidence angle $\varphi=34.7^ \circ$.}
\label{fig2-q-factor}
\end{figure}

In order to provide a coupling between the surface state and
incoming/outgoing radiation we consider a semi-infinite photonic
crystal structure containing only a finite number $N$ of the
surface rods of the reduced diameter $d=0.5D$. These rods define a
resonant cavity situated at the surface of the photonic crystal as
illustrated in the inset to Fig. \ref{fig2-q-factor} for the case
of $N=6$.

Our calculations are performed in the supercell geometry that
infinitely extends into the air region to the left and into the
bulk of the photonic crystal to the right (see Ref.
[\onlinecite{we-PRB-2005}] for details). In the transverse
direction (i.e. in the direction parallel to the surface) we
utilize the cyclic boundary condition. In our calculation the
supercell consists of $N+4$ unit cells in the transverse
direction. Because of the rapid decay of the field intensity into
the bulk, this choice is proven to be sufficient to ensure that
the neighboring supercells are well isolated from each other.

In order to calculate the $Q$ factor of the structure at hand, we
illuminate the surface by an incidence wave (that excites the
surface modes) and compute the intensity of the field
distribution. The $Q$ factor, defined as $Q$=2$\pi \omega
$*(stored energy)/(energy lost per cycle), then can be expressed
in the following form \cite{Sakoda,we-PRB-2005}:
\begin{equation}  \label{quality_factor}
Q=\omega \frac{\Omega }{4\int S_{in}dy},
\end{equation}%
where $\Omega _{TM}=\int [\varepsilon \varepsilon
_{0}|E_{z}|^{2}+\mu _{0}(|H_{x}|^{2}+|H_{y}|^{2})]dxdy$
characterizes the energy stored in the system (TM-polarization)
and the integral over $S_{in}$ is the incoming  energy flux. It
should be stressed the value of the $Q$ factor at resonance
depends on the coupling of the surface state modes with the
\textit{outgoing} radiation, and thus is independent of the
incidence angle of the incoming wave.

Note that the calculated field distribution includes the
contributions from both the surface mode (exited by the incident
light), and the incident and reflected waves. This leads to a
nearly constant off-resonance background in the dependence
$Q=Q(\omega)$ that is caused by the contribution of the incident
and reflected waves in the total field intensity in Eq.
(\ref{quality_factor}). To remove this background we calculate the
$Q$ factor of a structure without surface states. We choose this
structure as a semi-infinite photonic crystal with all identical
rods, which is known not to support surface modes
\cite{joannopoulos}. Then the obtained value is subtracted from
the calculated value of the $Q$ factor of the system under study
\cite{we-PRB-2005}.

Figure \ref{fig2-q-factor} shows a $Q$ factor of the resonant
cavity as a function of the frequency of the illuminating light.
In the given frequency interval there are three cavity modes with
the  $Q$ factors of the order of $\sim 10^{5}-10^{6}$. (Note that
above values might underestimate the actual theoretical $Q$
factors obtained within present 2D calculations, because even
finer frequency steps in the vicinity of the resonances are
required for better resolution of the $Q$ factors). The calculated
values of the $Q$ factors are comparable to those predicted and
achieved in defect mode bandgap cavities \cite{defect-mode,Noda}
and in whispering-gallery trajectories cavities
\cite{Slusher_1992,Gayral,we-JAP-2003, Vahala}. It is expected
however that in actual photonic structures realized typically in a
slab geometry, the $Q$ factor will be reduced due to the radiative
decay in the direction perpendicular to the plane of the photonic
crystal \cite{Noda} (which has not been accounted for in the
present 2D calculations).

Let us now analyze the positions of the calculated resonance
peaks. The structure at hand can be considered as a conventional
Fabry-Perot resonator whose resonant wavelengths are given by
$\lambda _{\alpha}=2\pi/k_{\alpha }$, with the wavevector
\begin{equation}
 k_{\alpha
}=\frac{\pi \alpha }{w},  \label{resonant_frequency}
\end{equation}%
where $\alpha $ is the mode number and $w$ is the width of the
cavity ($7a$ in the present case). For the photonic crystal under
study the surface state exists only in a limited frequency
interval, $0.33\lesssim \omega a /2\pi c\lesssim 0.42$ (see the
dispersion relation for the surface state in Fig.
\ref{fig1-disp-rel}). It follows from this dispersion relation
that all the modes given by Eq. (\ref{resonant_frequency}), except
$\alpha=5,6,7$ are situated outside this interval. An estimation
of the expected positions for the resonant peaks for the modes
$\alpha =5,6,7$ based on Eq. (\ref{resonant_frequency}) and on the
dispersion relation for the surface mode is shown in Fig.
\ref{fig2-q-factor} where the discrepancy between the expected and
calculated resonance frequencies does not exceed $0.5\%$.
\begin{figure}[htb!]
\begin{center}
\includegraphics[keepaspectratio,width=0.8\columnwidth]{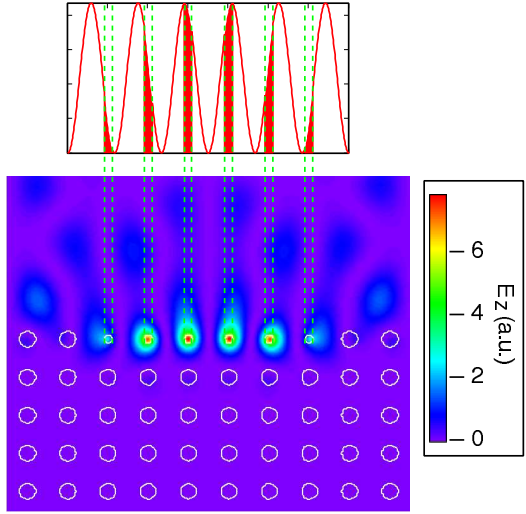}
\end{center}
\caption{(color online) Lower panel: Calculated intensity of the
$E_z$ component for the 6th mode of a resonant cavity shown in
Fig. \ref{fig2-q-factor}. Upper panel: Expected field intensity at
different rods is given by the overlap of the 6th eigenstate of
the cavity with the actual positions of the rods.}
\label{fig3-ez-ss}
\end{figure}

Figure \ref{fig3-ez-ss} illustrates the intensity of the $E_z$
component of the electromagnetic field for the resonance mode
$\alpha=6$. As expected, the field is localized in the cavity
inside the rods, and the intensity dies off very quickly both to
the open space and to the photonic crystal. The field intensity at
different rods in the cavity is expected to be determined by the
overlap of the $\alpha$th eigenstate of the Fabry-Perot resonator
with the actual positions of the rods in the cavity. This overlap
for the 6th mode is shown in Fig. \ref{fig3-ez-ss}, which agrees
very well with the actual calculated intensity distribution
pattern. Note that all the results reported in this Letter
correspond to the case of $N=6$ rods (see inset to Fig.
\ref{fig2-q-factor}). We performed calculations for different
numbers of rods $N=5-11$ and we always find an excellent agreement
between the calculated and expected resonant frequencies as well
as between the intensity distributions.

Recently, it has been demonstrated that a surface of a photonic
crystal supporting a leaking surface mode can serve as a kind of
antenna to beam the emitted light in a single direction, which may
be used for integration of the photonic crystals with conventional
fiber optics devices or lenses \cite{Moreno, Kramper}. The
proposed surface mode photonic bandgap cavities, thanks to their
unusual properties where the electromagnetic field resides
directly on the surface, can also lead to new applications and
novel integration schemes. These might include utilization of the
surface state cavities for lasing and sensing applications as well
as integration of the photonic crystal devices with e.g.
whispering-gallery-trajectory microlasers that can be directly
coupled to the surface mode cavities.

To conclude, we propose and analyze a novel high-$Q$ cavity based
on a surface state of a photonic crystal. The resonant cavity is
localized directly on the surface row of the photonic crystal.
This unusual property of the surface mode photonic bandgap
cavities might open up new possibilities for design and
integration of novel photonic devices for lasing, sensing and
filtering applications.

\end{document}